\documentclass[aps,prb,twocolumn,showpacs]{revtex4}
\usepackage{graphicx}
\usepackage{dcolumn}
\usepackage{color}
\usepackage{mathrsfs}
\usepackage[breaklinks,colorlinks = true,linkcolor = blue,urlcolor=blue,citecolor=blue]{hyperref}
\usepackage{amsmath}
\usepackage{soul}

\begin{document}
\title{Interplay of Spin and Lattice Degrees of Freedom in the Frustrated Antiferromagnet CdCr$_2$O$_4$: High-field and Temperature Induced Anomalies of the Elastic Constants}

\author{Subhro Bhattacharjee$^{1,2}$}
\email{subhro@physics.utoronto.ca}
\altaffiliation[Present address: ]{Department of Physics, University of Toronto, Ontario M5S 1A7, Canada; Department of Physics \& Astronomy, McMaster University, Hamilton, Ontario L8S 4M1, Canada.}

\author{S. Zherlitsyn$^3$}

\author{O. Chiatti$^3$}
\altaffiliation[Present address: ]{London Centre for Nanotechnology, University College London, 17-19 Gordon Street, London WC1H 0AH, United Kingdom.}

\author{A. Sytcheva$^3$}

\author{J. Wosnitza$^3$}

\author{R. Moessner$^1$}

\author{M.E. Zhitomirsky$^4$}

\author{P. Lemmens$^5$}

\author{V. Tsurkan$^{6,7}$}

\author{A. Loidl$^6$}

\affiliation{
$^1$ Max-Planck-Institut f\"ur Physik komplexer Systeme, Dresden D-01187, Germany\\
$^2$ Department of Physics, Indian Institute of Science, Bangalore-560012, India\\
$^3$ Hochfeld-Magnetlabor Dresden, Helmholtz-Zentrum Dresden-Rossendorf, Dresden D-01314, Germany\\
$^4$ SPSMS, UMR-E9001 CEA-INAC/UJF, 17 rue des Martyrs, 38054 Grenoble Cedex 9, France\\
$^5$ IPKM, Technische Universit\"{a}t Braunschweig, Braunschweig D-38106, Germany\\
$^6$ Universit\"{a}t Augsburg, Augsburg D-86159, Germany\\
$^7$ Institute of Applied Physics, Academy of Sciences of Moldova, Chisinau MD-2028, Republic of Moldova}
\begin{abstract}
{{Temperature and magnetic field studies of the elastic constants of the chromium spinel CdCr$_2$O$_4$ show pronounced anomalies related to strong spin-phonon coupling in this frustrated antiferromagnet. A detailed comparison of the longitudinal acoustic mode propagating along the [111] direction with a theory based on an exchange-striction mechanism leads to an estimate of the strength of the magnetoelastic interaction. The derived spin-phonon coupling constant is in good agreement with previous determinations based on infrared absorption. Further insight is gained from intermediate and high magnetic field experiments in the field regime of the magnetization plateau. The role of the antisymmetric Dzyaloshinskii-Moriya interaction is discussed.}}
\end{abstract}

\maketitle
\section{Introduction}
Highly frustrated antiferromagnets (AFs) are of great interest due to their potential to realize various unconventional phases, even in the classical limit. An example in three dimensions is the much studied pyrochlore lattice, which consists of corner-sharing tetrahedra. On the theoretical side, it was noticed quite early that the classical Heisenberg AF with nearest-neighbour interactions on a pyrochlore lattice has a very unconventional  ground state, often referred to as a {\it{classical spin liquid}} or a {\it {cooperative magnet}}{\cite{1979_villain,1998_moessner}} that is characterized by power-law (dipolar) spin correlations.{\cite{2004_isakov,2005_henley}} Various perturbations in such a highly correlated paramagnet are known to have a drastic effect on the ground state. The discovery of several materials of this kind has led to a strong interest in magnets with a pyrochlore lattice.

A classic example of a pyrochlore Heisenberg AF is the spinel compound CdCr$_2$O$_4$. Here, the Cr$^{3+}$ ions with spin $S=3/2$ form the pyrochlore lattice. Although the Curie-Weiss temperature is $\Theta_{CW} \approx - 70$ K in this compound, the AF ordering sets in only at $T_{N}$ = 7.8 K, indicating a high level of magnetic frustration.{\cite{2005_chung}} The intermediate regime, also referred to as a cooperative paramagnet, already exhibits strong but short-range spin correlations.\cite{2005_ueda} The magnetic ordering at $T_{N}$ is accompanied by a structural change with loss of inversion symmetry from the cubic ($Fd\overline{3}m$) to the tetragonal ($I4_1/amd$) (Ref. {\onlinecite{2010_matsuda}}) structure. This first order {\it {magneto-structural}} phase transformation is caused by a {\it{spin Jahn-Teller effect}}, whereby the spins relieve their frustration by distorting the crystal.\cite{2002_tchernyshyov} Neutron-scattering studies \cite{2005_chung,2007_matsuda} show a long-pitched incommensurate coplanar spiral spin configuration below $T_{N}$ [ordering wave vector ${\bf Q}=2\pi(0,\delta,1)$, $\delta=0.09$]. This, in turn, is attributed to weak Dzyaloshinskii-Moriya (DM) interactions\cite{2006_chern} that are present in this compound and assert themselves in the magnetically ordered phase. Electron Spin Resonance (ESR) measurements suggest {\cite{2006_kimura}} that, in a magnetic field ($H\approx 5.7$ T), the spiral transforms to a four-sublattice canted structure. Pulsed-field magnetization data exhibit a broad plateau between 28 and about 60 T, \cite{2005_ueda, 2008_kojima} which corresponds to 1/2 of the saturation magnetization of the Cr$^{3+}$ magnetic moments. The fully polarized state is achieved above 90 T. \cite{2008_kojima} The anisotropy of the magnetic exchange interactions otherwise appears to be negligibly small in this compound. \cite{2005_ueda} All experiments clearly indicate the importance of spin-lattice coupling in this compound and, indeed, in the whole family of chromium spinels (ACr$_2$O$_4$, A = Zn, Cd, Hg).

Ultrasound investigation is a powerful method to probe spin-lattice interactions and possible lattice instabilities \cite{2005_luthi}. Indeed, it is an ideal experimental tool to investigate the chromium spinels as the {\em magnetodistortive} transition is precisely about the interplay of spin and lattice degrees of freedom. We have performed measurements of the relative change of the sound velocity in CdCr$_2$O$_4$ using a phase-sensitive detection technique, based on a standard pulse-echo method with a setup as described in more detail in Refs. [\onlinecite{2005_luthi, 2001_Wolf}]. This technique is available as well for non-destructive pulsed magnetic fields, extending the parameter range for measuring the sound velocity and sound attenuation up to very high magnetic fields. The measurement accuracy for a relative change of sound velocity is of the order of 10$^{-6}$ for static-field measurements and 10$^{-5}$ for the pulsed-field experiments. The sound velocity $v$ is related to the elastic constant $c=\rho v^2$, where $\rho$ is the mass density of the crystal. In this paper, we report on the relative change of the sound velocity for a longitudinal acoustic, $c_{L}$, mode propagating along the [111] direction of CdCr$_2$O$_4$ (of the undistorted cubic crystal) both in static fields up to 18 T and pulsed fields up to 63 T. Both magnets were equipped with ${^4}$He-flow cryostats. Wide frequency-range piezoelectric film transducers glued on the parallel surfaces of the sample with a two-component epoxy were used for the excitation and detection of the longitudinal sound waves.

 The acoustic mode under consideration corresponds to $c_{L} = 1/3(c_{11}+2c_{12}+4c_{44})$, with $c_{ij}$ being the elastic constants for a cubic crystal (we have used the familiar Voigt notations). The magnetic field was applied along the [111] direction, i.e., parallel to the wave vector ${\bf k}$ and the polarization ${\bf u}$ of the ultrasound wave. The magnetization has been measured using a commercial SQUID (superconducting quantum interference device) magnetometer equipped with a 7 T magnet. A high-quality single crystal of CdCr$_2$O$_4$ was obtained by spontaneous crystallization from Bi$_2$O$_3$-V$_2$O$_5$ flux. The sample thickness along the [111] direction is 0.83 mm.

The sound velocity shows anomalous dependences as a function of both temperature and magnetic field. We provide a theoretical framework to understand these effects on the basis of spin-lattice coupling. Our paper reveals a fascinating interplay of spin and lattice degrees of freedom in this highly frustrated magnet.

The paper is organized as follows. In Sec. \ref{sec_micro_ham}, we introduce the microscopic Hamiltonian that captures the spin-phonon interactions in CdCr$_2$O$_4$ and this forms the basis of our analysis. The effect of the spin-lattice interaction on the sound velocity is divided up according to the magnetic phases. We start by discussing the behavior in the paramagnetic phase in Sec. \ref{sec_para_phase}. Then, we briefly, discuss the features of the magnetoelastic transition in Sec. \ref{sec_transition}. This is followed by the discussion of the low-temperature phase in Sec. \ref{sec_low_t} and the effect of the magnetic field in Sec. \ref{sec_mag_fld}. Finally, we summarize our conclusions in Sec. \ref{sec_conclusion}. The details of various calculations are given in Appendices \ref{appen} and \ref{appen2}.
\section{The Spin-Phonon Hamiltonian}

\label{sec_micro_ham}

In CdCr$_2$O$_4$, the CrO$_6$ octahedra build an edge-sharing network with Cr$^{3+}$  ions forming a pyrochlore lattice. The octahedral symmetry of the crystal field splits the five Cr d-orbitals and lowers the energy of the three t$_{2g}$ orbitals compared to the two e$_g$ orbitals. Strong Hund's coupling aligns the three electrons in the t$_{2g}$ orbital leaving a net spin of $S$ = 3/2 on each Cr$^{3+}$ ion. The orbital part of the electron wave function forms a singlet, and the orbital degrees of freedom are effectively quenched. At low temperatures, a spin-only Heisenberg Hamiltonian suitably describes the system. The non-magnetic Cd ions control the Cr-Cr distance and thereby the value of the exchange strength. Further-neighbour exchanges are weak owing to the arrangements of different relevant overlapping orbitals. {\cite{2008_ueda, 2010_bhattacharjee}} Thus, the principal part of the spin Hamiltonian given by
\begin{eqnarray}
\label{04_heisen_ham}
H'_{\mathrm{sp}}=\sum_{\langle ij\rangle} J_{ij}{\bf S}_i\cdot {\bf S}_j-\sum_i{\bf h}\cdot {\bf S}_i
\end{eqnarray}
encodes the antiferromagnetic $(J_{ij}>0$) exchange between nearest neighbours $\langle ij\rangle$. The second term denotes the usual Zeeman coupling to an external magnetic field ${\bf H}$, with ${\bf h}=g\mu_B {\bf H}$ ($g$ and $\mu_B$ are the gyromagnetic ratio and the Bohr magneton, respectively).

In the absence of orbital degrees of freedom, the magnetoelastic coupling is mediated by the dependence of $J_{ij}$ on the position of the Cr spins. (Throughout our calculations, we treat the spins as classical.) By expanding $J_{ij}$ in Eq. (\ref{04_heisen_ham}) around the equilibrium positions of the Cr ions to harmonic order, we get
\begin{eqnarray}
J_{ij}=J_0+\frac{\partial J_{ij}}{\partial{\bf R}_i}\cdot {\bf R}_{ij}+\frac{1}{2}{\bf R}_{ij}\cdot\frac{\partial^2J_{ij}}{\partial{\bf R}_{ij}^2}\cdot {\bf R}_{ij}.
\label{spn_ph}
\end{eqnarray}
$J_0$ is the equilibrium exchange coupling and ${\bf R}_{ij}={\bf R}_i-{\bf R}_{j}$, where ${\bf R}_i$ is the displacement of $i^{th}$ Cr ion from its equilibrium position. For the full spin-phonon Hamiltonian, we must introduce phonons. Writing bosonic creation and annihilation operators, $b_{\bf k}$ and $b^\dagger_{\bf k}$, for the phonons, we obtain
\begin{eqnarray}
\label{h_int}
H&=&H_{\mathrm{ph}}+H_{\mathrm{sp}}+H_{\mathrm{ph-sp}},
\end{eqnarray}
with
\begin{subequations}
\begin{eqnarray}
H_{\mathrm{ph}}&=&\hbar\sum_{\bf k}\omega^0_{\bf k}b^\dagger_{\bf k}b_{\bf k},
\end{eqnarray}
\begin{eqnarray}
\label{spin_hamiltonian}
H_{\mathrm{sp}}&=&J_0\sum_{\langle ij\rangle}{\bf S}_i\cdot{\bf S}_j-\sum_i{\bf h}\cdot{\bf S}_i,
\end{eqnarray}
\begin{eqnarray}
H_{\mathrm{ph-sp}}&=&H_1+H_2.
\end{eqnarray}
\end{subequations}
We are interested in the longitudinal phonon mode with $\omega^0_{\bf k}$ and ${\bf e_k}$ as the bare frequency and the polarization, respectively. For long wavelengths, $\omega^0_{\bf k}=v^0_{{\bf k}} k$, where $v^0_{{\bf k}}$ is the bare sound velocity in the direction ${\bf k}$. $H_1$ and $H_2$, respectively, are the spin-phonon interactions that arise from the first- and second-order terms of Eq. (\ref{spn_ph}),
\begin{eqnarray}
H_1=\sum_{\bf k}U^{(1)}_{\bf k}A_{\bf k},\ \ \ \ H_2=\frac{1}{2}\sum_{\bf k k'}U^{(2)}_{\bf k k'}A_{\bf k}A_{-{\bf k'}},
\end{eqnarray}
where $A_{\bf k}=b_{\bf k}+b^{\dagger}_{-{\bf k}}$ and
\begin{subequations}
\begin{eqnarray}
\label{potential1}
\nonumber
U^{(1)}_{\bf k}=&&\sqrt{\frac{\hbar}{2 M N\omega_{\bf k}^0}}\sum_{\langle ij\rangle}{\bf S}_i\cdot{\bf S}_j\left( e^{\imath{\bf k}\cdot {\bf r}_i}-e^{\imath{\bf k}\cdot{\bf r}_j}\right)\\
&&\times\left(\frac{\partial J_{ij}}{\partial{\bf R}_i}\cdot {\bf e}_{\bf k}\right),
\end{eqnarray}
\begin{eqnarray}
\label{potential2}
\nonumber
U^{(2)}_{\bf k k'}=&&\frac{\hbar}{2\ M N}\frac{1}{\sqrt{\omega_{\bf k}^0\omega_{\bf k'}^0}}\sum_{\langle ij\rangle}{\bf S}_i\cdot{\bf S}_j\left( e^{-\imath{\bf k'}\cdot{\bf r}_i}-e^{-\imath{\bf k'}\cdot{\bf r}_j}\right)\\
&&\times\left({\bf e}_{-\bf k'}\cdot\frac{\partial^2 J_{ij}}{\partial{\bf R}_{ij}^2}\cdot{\bf e}_{{\bf k}}\right)\left( e^{\imath{\bf k}\cdot {\bf r}_i}-e^{\imath{\bf k}\cdot{\bf r}_j}\right).
\end{eqnarray}
\end{subequations}
Here, $M$ is the mass of the chromium ion (note that the same symbol will be used for magnetization later; the context will clarify the meaning) and $N$ is the total number of chromium ions. In addition to nearest neighbor exchange, CdCr$_2$O$_4$ has weak further-neighbor exchanges and DM interactions. While the effect of the former are expected to be negligible, the DM interactions play a crucial role in the low-temperature ordered state of CdCr$_2$O$_4$,{\cite{2006_chern}} as we will see later.

Useful simplifications occur in the Hamiltonian given by Eq. (\ref{h_int}) on noticing that the direct exchange dominates the magnetic interaction between neighbouring Cr spins.\cite{2008_ueda} Thus, we have $J_{ij}=J_0e^{-\alpha{R_{ij}}}$, where $\alpha$ now determines the strength of the magnetoelastic interactions.

The effect of the spin-lattice coupling on the spins in cubic spinels in similar context was discussed by Tchernyshyov {\it {et al.}}\cite{2002_tchernyshyov} They considered the Hamiltonian given by Eq. (\ref{04_heisen_ham}), expanded $J_{ij}$ to linear order ($\partial J_{ij}/\partial{\bf R}_i$), and showed that the spin degeneracy is lifted through a Jahn-Teller-type mechanism selecting a collinearly spin-ordered state, and at the same time, distorting the lattice.  On including the second-order term $\partial^2 J_{ij}/\partial{\bf R}_{ij}^2 $, the potential energy for the bond $(ij)$ is given by
\begin{eqnarray}
\tilde{V}_{ij}=\frac{1}{2}K\delta_{ij}^2+J^\prime {\bf S_i\cdot S_j} \delta_{ij}+\frac{1}{2}J^{\prime\prime}{\bf S_i\cdot S_j}\delta_{ij}^2,
\end{eqnarray}
where $K$ is the stiffness constant of the bond and $J',J''$ are the first- and second-order derivatives of $J_{ij}$ as indicated in Eq. ({\ref{spn_ph}}). Integrating out the displacement variables ($\delta_{ij}$) within the independent-bond approximation, we have:
\begin{eqnarray}
\tilde{V}_{ij}=-\frac{(J^\prime{\bf S}_i\cdot{\bf S}_j)^2}{2(K+J^{\prime\prime}{\bf S}_i\cdot {\bf S}_j)}.
\label{second_effect}
\end{eqnarray}
Thus, in the limit $K\gg J^\prime,J^{\prime\prime} $, it is clear from Eq. (\ref{second_effect}) that the second-order term does not destabilize the collinear spin ordering but renormalizes the different order parameters at sub-leading order, leaving the general structure of the phase diagram intact. (Similar results are obtained in a more detailed analysis.{\cite{2010_bhattacharjee}}) It is very important to note that this is not the case for the renormalization of the phonon spectrum.

The effect of the spin-phonon interaction on the phonon spectrum is effectively studied using perturbation theories suited to specific regimes, namely, the paramagnetic and the magnetically ordered regimes. We address these regimes in the following sections.
\section{The Paramagnetic Phase}

\label{sec_para_phase}

In the absence of spin-phonon coupling, in the temperature regime $T_N<T< \vert\Theta_{CW}\vert$, the spin correlations tend toward a form that falls off with a fixed integer power of distance and a characteristic dipolar angle dependence. These correlations lead to a non trivial structure factor in the neutron scattering in this {\it{cooperative paramagnetic}} phase. {\cite{1979_villain}} Furthermore, in this phase, the spin dynamics has an unusual behavior, {\cite{1998_moessner2, 2009_conlon}} characterized by an emergent universal (independent of exchange strength) timescale, $\tau_{s}\sim \hbar/ck_BT$, $c=O(1)$, that controls the long-time dynamics for generic wave vectors. It is known that such behavior survives in the presence of weak spin-lattice coupling. However, the behavior of the phonons has not been studied so far.

\begin{figure}
\begin{center}
\vspace{-0.5cm}
\includegraphics[scale=0.25]{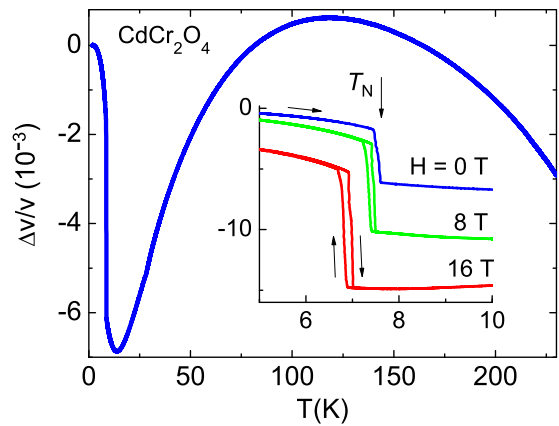}
\end{center}
\vspace{-0.5cm}
\caption{(Color online) Change of the sound velocity vs temperature at $H$ = 0 for the $c_{L}$ mode in CdCr$_2$O$_4$ for an ultrasound frequency of 107 MHz. The inset shows the region around $T_{N}$ in more detail. Data for 8 T (middle curve) and 16 T (lower curve) applied along the [111] direction are shown as well. For the non zero magnetic fields, the temperature sweeps up and down are shown. Notice that the transition temperature is reduced on increasing the field. All data curves are arbitrarily shifted along the y-axis for clarity.}
\label{compare1}
\end{figure}

Figure [\ref{compare1}] shows the temperature dependence of the sound velocity of the $c_{L}$ mode measured below 240 K at zero magnetic field. It exhibits a softening below 120 K followed by a minimum at approximately 13 K and a jump like anomaly at $T_{N}$, which is accompanied by a small hysteresis signaling the magneto-structural transition. The jump shifts to lower temperatures in an applied magnetic field [see Fig. \ref{compare1} (inset) and also Fig. \ref{field_fig}]. This lowering with increasing magnetic field is understood from the Clausius-Clapeyron equation for the magnetic systems: $\frac{dT_c}{dH}=-\frac{\Delta M}{\Delta S}$. Magnetization measurements (Fig. \ref{compare3}) show $\Delta M < 0$ as one goes from the high-temperature to the low-temperature phase. Also, $\Delta S< 0$ as the system moves into an ordered state from the frustrated paramagnet. Hence, $\frac{dT_c}{dH}< 0$ as seen in experiment. The gradual decrease of the sound velocity occurs at temperatures corresponding to the paramagnetic state of the spins. Note that the infrared reflectivity spectra exhibit a phonon softening in the same temperature range.\cite{2007_rudolf, 2008_aguiliar, 2009_kant}

Using the Hamiltonian introduced in Eq. (\ref{h_int}) we can make quantitative estimates of the phonon self-energy correction due to coupling with the spins. In the paramagnetic phase, by comparing the time scales of the acoustic phonons and the spins, we find that the spin dynamics is fast, and, hence, the spins may be integrated out. Here, it is worthwhile to note that the opposite limit is obtained in case of the optical phonons, which is relevant for the above-mentioned infrared measurements. {\cite{1988_lockwood,1970_baltensperger}} Integrating out the spins results in an effective interaction among the acoustic phonons given by the effective Hamiltonian,
\begin{eqnarray}
\label{eff_act}
\mathcal{H}_{\mathrm{eff}}=\sum_{\bf k}\omega^0_{k}b^\dagger_{\bf k}b_{\bf k}+\frac{1}{2}\sum_{\bf k,k'}V_{\bf k k'} A_{\bf k}A_{-\bf k'},
\end{eqnarray}
where
\begin{eqnarray}
V_{\bf k k'}=\langle U^{(2)}_{\bf k k'}\rangle -\beta \langle\langle U^{(1)}_{\bf k}U^{(1)}_{-\bf k'}\rangle\rangle.
\label{eff_potential}
\end{eqnarray}
Here $\langle\cdots\rangle$ denotes thermal averaging over the spins [with respect to the unperturbed Heisenberg Hamiltonian $H_{\mathrm{sp}}$ given by Eq. (\ref{spin_hamiltonian})] and $\langle\langle\cdots\rangle\rangle$ stands for the connected correlator: $\langle\langle AB\rangle\rangle=\langle AB\rangle-\langle A\rangle\langle B\rangle$. Also $\beta=1/K_BT$, where $K_B$ is the Boltzmann constant. Since the timescale for the energy exchange between the phonons and the spins is inversely proportional to the spin-phonon scattering rate, this perturbation calculation is valid only when the time scale associated with the energy exchange between the spins and the {\it acoustic} phonons is the longest timescale in the problem. At present, although there is no estimate of this timescale, our results provide some justification for this assumption {\it a posteriori}. {\cite{footnote}}

We start our calculation by writing the Matsubara-Green's function for the phonons, $\mathcal{G}^\beta({\bf q},\tau-\tau')=-\langle \mathcal{T}\{A_{\bf q}(\tau)A_{-\bf q}(\tau')\}\rangle$, and find the contribution to the phonon self-energy to the lowest order in spin-phonon coupling. The dressed phonon propagator is given by
\begin{eqnarray}
\label{matsubara_green}
\mathcal{G}^\beta({\bf q},i\Omega_n)=\frac{2\omega^0_{\bf q}}{(\imath\Omega_n)^2-(\omega^0_{\bf q})^2-2\omega^0_{\bf q}\Sigma(\imath\Omega_n,{\bf q})},
\end{eqnarray}
where ${\bf q}$ is the wave vector, $\Omega_n=2\pi n/\beta$ is the bosonic Matsubara frequency and $\Sigma(\imath\Omega_n,{\bf q})$ is the phonon self-energy, which is given by
\begin{eqnarray}
\label{self_energy}
\Sigma(\imath\Omega_n,{\bf q})=V_{\bf q q},
\end{eqnarray}
where $V_{\bf q q}$ is given by Eq. (\ref{eff_potential}). From this we find the leading-order change of the phonon frequency as
\begin{eqnarray}
\label{real_self_energy}
\nonumber
\Delta\omega_{\bf q}&=&\omega^0_{\bf q}\left(\sqrt{1+\frac{2}{\omega^0_{\bf q}}\mathrm{Re}\left[\Sigma(\imath\Omega_n,{\bf q})\vert_{\imath\Omega_n\rightarrow\omega_{\bf q}+\imath\eta}\right]}-1\right)\\
&\approx&\mathrm{Re}\left[\Sigma(\imath\Omega_n,{\bf q})\vert_{\imath\Omega_n\rightarrow\omega_{\bf q}+\imath\eta}\right].
\end{eqnarray}
Hence, the fractional change in the velocity of sound is given by 
\begin{eqnarray}
\label{delta_v_expression}
\frac{\Delta v}{v}=\lim_{q\rightarrow 0}\frac{\Delta \omega_{\bf q}}{\omega^0_{\bf q}},
\end{eqnarray}
As shown in Appendix \ref{appen1}, for the $[111]$ direction of the sound, Eq. (\ref{delta_v_expression}) becomes
\begin{eqnarray}
\nonumber
\frac{\Delta v}{v}=\frac{1}{9 M (v^0_{\hat{\bf q}})^2}&&\left[\left(\frac{\delta^2}{J_0}\frac{\partial^2 J}{\partial \delta^2}\right)\mathcal{E}^{\mathrm{spin}}\right.\\
&&\left.-\left(\frac{\delta}{J_0}\frac{\partial J}{\partial\delta}\right)^2C^{\mathrm{spin}}T\right],
\label{semifinal_exp}
\end{eqnarray}
where $C^{\mathrm{spin}}$ and $\mathcal{E}^{\mathrm{spin}}$ are the unperturbed magnetic specific heat and average energy, respectively, of the single spin for a nearest-neighbor Heisenberg AF on a pyrochlore lattice [Eq. (\ref{spin_hamiltonian}) with ${\bf h}=0$]. Inclusion of nonlinear terms into the phonon Hamiltonian (\ref{eff_act}) (as well as probing sound velocity in directions other than $[111]$ [see Appendix \ref{appen1}]) may modify the coefficients in front of $\mathcal{E}^{\mathrm{spin}}$ and $C^{\mathrm{spin}}T$ in the above expression, but we neglect the corresponding effects. In the cooperative paramagnetic phase of the classical Heisenberg AF it was shown {\cite{1999_moessner}} that the energy and specific heat can be calculated quite accurately using a single tetrahedron approximation. Thus, we have [using direct exchange for the spin-spin coupling ($J=J_0e^{-\alpha \delta}$)]
\begin{figure}
\centering
\includegraphics[angle=0,scale=0.27]{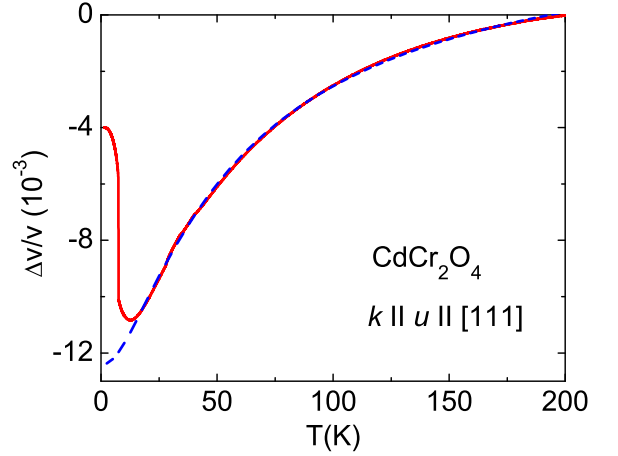}
\caption{(Color online) The best fit (dashed line) of the zero magnetic field experimental data (solid line) in the paramagnetic regime and determination of $\alpha$. The non harmonic part present in the experimental data (in Fig. \ref{compare1}, for ${\bf H}=0$) has been subtracted (using Ref. [\onlinecite{1970_varshni}], see text for details). The data have been fitted in the regime of $T=20-200$ K using Eq. (\ref{final_exp}). The variation in $\alpha$ (calculated from these fitting) is due to the uncertainty in the value of $J$ ($J$ = 0.80 -1.00 meV). Note the good quality of the fit down to $T\ll\vert\Theta_{CW}\vert$.}
\label{fitting_fit_t}
\end{figure}
\begin{eqnarray}
\label{final_exp}
\frac{\Delta v}{v}=\frac{\alpha^2\delta^2}{18 M (v^0_{\hat{\bf q}})^2}\left(\mathcal{E}^{\mathrm{Tet}}-TC^{\mathrm{Tet}}\right),
\end{eqnarray}
where $\mathcal{E}^{\mathrm{Tet}}$ and $C^{\mathrm{Tet}}$ are the energy and specific heat per tetrahedron. To compare with experiments, it is useful to factor out $J_0S^2$ from the expressions of $\mathcal{E}^{\mathrm{Tet}}-T C^{\mathrm{Tet}}$ by expressing them as a function $T/(J_0S^2)=T_0$. Thus Eq. (\ref{final_exp}) becomes
\begin{eqnarray}
\label{scaled_final_exp}
\frac{\Delta v}{v}=\frac{\alpha^2\delta^2J_0S^2}{18 M (v^0_{\hat{\bf q}})^2}\left(\overline{\mathcal{E}}^{\mathrm{Tet}}-T_0\overline{C}^{\mathrm{Tet}}\right),
\end{eqnarray}
where $\overline{\mathcal{E}}^{\mathrm{Tet}}$ and $\overline{C}^{\mathrm{Tet}}$ are functions of the single parameter $T_0$, which can be calculated exactly in the absence of a magnetic field.\ {\cite{1999_moessner}} The explicit expressions are given in Appendix \ref{appen2}. The factor 1/2 is multiplied in Eq. (\ref{final_exp}) because each spin is shared by two tetrahedra. 

In the expression given by Eq. (\ref{scaled_final_exp}) the numerical values of all parameters, except $\alpha$  are independently known. Hence, we can use a single-parameter fit describing the experimental data to get an estimate of $\alpha$. However, there is a non-harmonic contribution to the temperature dependence of the sound velocity clearly seen in Fig. \ref{compare1} at $T$ above 120 K. This contribution is superimposed on the sound velocity renormalization arising from the spin-phonon coupling. Applying a well-established procedure, we have subtracted the non harmonic contribution from the experimental velocity data of Fig. \ref{compare1} using an empirical equation from Ref. \onlinecite{1970_varshni}. The fit of the corresponding experimental data is shown in Fig. \ref{fitting_fit_t} for $\alpha={13.7\pm 0.6\ {\rm \AA}^{-1}}$. (The error bar is mainly due to the uncertainty in the determination of the exchange coupling $J_0$.) Aguilar {\em et al.} {\cite{2008_aguiliar}} and Kant {\em et al.} {\cite{2009_kant}} measured the shift of the infrared-active optical phonon. From their optical-phonon data, we find $\alpha=10.97\pm 1.24~{\rm \AA}^{-1}$. Thus, the coupling constants derived from optical and acoustic phonons are quite close to each other.
\section{The Magneto-Elastic Transition}
\label{sec_transition}
\begin{figure}
\centering
\includegraphics[scale=0.27]{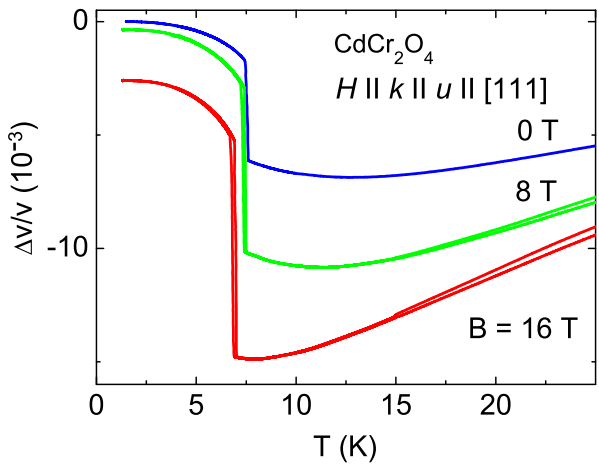}
\caption{(Color online) The sound velocity vs. temperature at different fields. Notice that the upturn is suppressed on increasing the field (see text for details). The data curves are arbitrarily shifted along the y-axis for clarity.}
\label{field_fig}
\end{figure}

The hysteresis in the sound velocity at $T_{N}$ confirms the first-order type of the phase transition, previously also observed in magnetic-susceptibility measurements. \cite{2005_ueda} The sound attenuation (not shown) increases on lowering the temperature and shows a peak like anomaly at $T_{N}$. \cite{2010_zherlitsyn} The strong anomalies in the acoustic properties at the transition reflect the crucial role of the spin-lattice coupling in this compound. A particularly interesting feature is the smooth upturn (from about $T=13$ K) in the sound velocity as a precursor to the actual jump at the magnetoelastic transition (see Fig. \ref{compare1}). Similar anomalous features have also been seen in specific-heat measurements.{\cite{2009_kant}} This upturn is suppressed with increasing magnetic field (Fig. [\ref{field_fig}]). At present there is no clear understanding of its origin. The fact that the upturn is suppressed by the magnetic field indicates that it results from spin fluctuations. This suggests the presence of near-critical modes in the vicinity of the magnetoelastic transition that are ultimately cut off at some length scale leading to the actual first-order transition.

A Landau free-energy analysis based on the interaction of the magnetic modes and strain fields{\cite{2010_bhattacharjee}} qualitatively accounts for the actual jump of the sound velocity (the jump being proportional to the magnitude of the order parameter on the low-temperature side). However, such a phenomenological theory is highly qualitative due to the large number of terms allowed by various symmetries.
\section{The Low-Temperature Ordered Phase}
\label{sec_low_t}
\begin{figure}
\centering
\includegraphics[scale=0.27]{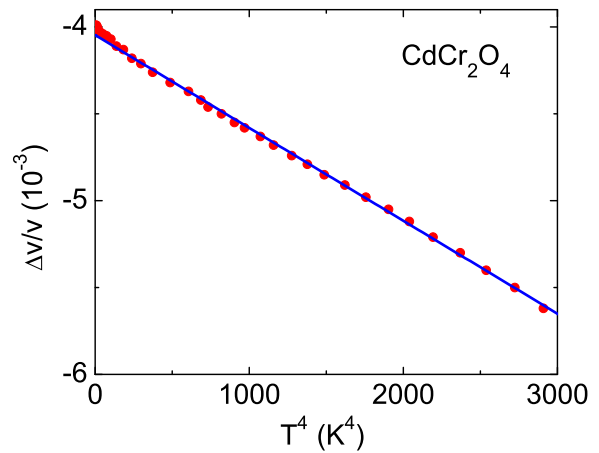}
\caption{(Color online) Best fit (solid line) of the low-temperature data (symbols) (in zero magnetic field) using Eq. (\ref{low_t}). The data clearly show an $\sim T^4$ behavior as expected from the theory of scattering of magnons and phonons.}
\label{compare5}
\end{figure}

In the low-temperature state, the lattice is distorted, and the spins are ordered in a long-pitched spiral. Owing to the lattice distortion, the erstwhile $[111]$ direction is no longer a symmetry axis. Thus, it is difficult to extract the magnetoelastic coupling constant(s) in this region and to check their consistency with that obtained in the paramagnetic regime.

However, in this regime, an analysis based on the magnon-phonon scattering in a collinear AF gives insights into the temperature dependence of the sound velocity. We can no longer integrate out the spins due to the presence of the low-frequency long-wavelength spin waves characteristic to the broken spin-rotation symmetry phase. Thus, essentially, we have a problem of magnon-phonon scattering, and we wish to calculate the renormalization of the phonon energy due to these scatterings. Hence, we start again with the Hamiltonian in Eq. (\ref{04_heisen_ham}) and use the Holstein-Primakoff (HP) approximation for the spin-dependent potentials $U^{(1)}_{\bf k}$ and $U^{(2)}_{\bf k,k'}$ [Eqs. (\ref{potential1}) and (\ref{potential2})].  Then, using diagrammatic perturbation theory as before, we calculate the renormalized phonon propagator [as in Eq. (\ref{matsubara_green})]. The calculation of the phonon self energy is a problem of evaluating the various magnon-phonon diagrams.{\cite{2007_cheng,1995_dirk,2001_woods}} After some rather tedious but ultimately straight forward diagrammatic calculation [outlined in Appendix \ref{appen1b}], we obtain the phonon self-energy. Then, using Eqs. (\ref{real_self_energy}) and (\ref{delta_v_expression}), we compute the fractional change in the sound velocity (to the leading order at ${\bf H}=0$)
\begin{eqnarray}
\label{low_t}
\frac{\Delta v}{v}= (c + K T^4).
\end{eqnarray}
The coefficients $c$ and $K$ depend on the form of the lattice and direction of the sound wave, in general, and are hard to determine theoretically, particularly for a distorted crystal. However, on general grounds we can argue that $c<0$. This is essentially the contribution arising from $U^{(2)}_{\bf k,k'}$ in Eq. (\ref{potential2}) with the spins in their classical ground state. In this case, it is easy to see that $c<0$ which indicates that the phonon mode will be softened at $T=0$ due to the coupling with the spins.  Figure \ref{compare5} shows such a fit, which compares fairly well with the experimental data. [A similar calculation for a ferromagnet predicts that $\frac{\Delta v}{v}\propto (c + K T^2)$.]
\section{\textbf{Measurements} in non zero \textbf{Magnetic Fields}}
\label{sec_mag_fld}

\begin{figure}
\begin{center}
\vspace{0.25cm}
\includegraphics[width=0.9\linewidth, keepaspectratio]{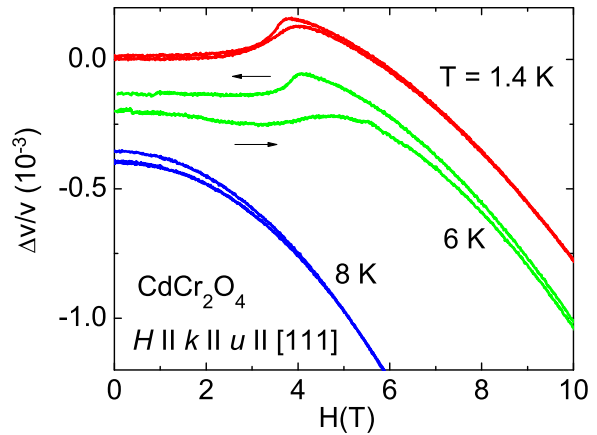}
\end{center}
\vspace{-0.5cm}
\caption{(Color online) Change of the sound velocity of the $c_{L}$ acoustic mode in CdCr$_2$O$_4$ vs magnetic field measured at different temperatures. The ultrasound frequency was 107 MHz. The arrows indicate the field-sweep directions. The experimental geometry is $H \parallel k \parallel u \parallel [111]$. The curves obtained at different temperatures are arbitrarily shifted along the y-axis for clarity.}
\label{compare2}
\end{figure}

\begin{figure}
\centering
\includegraphics[angle=0,scale=0.25]{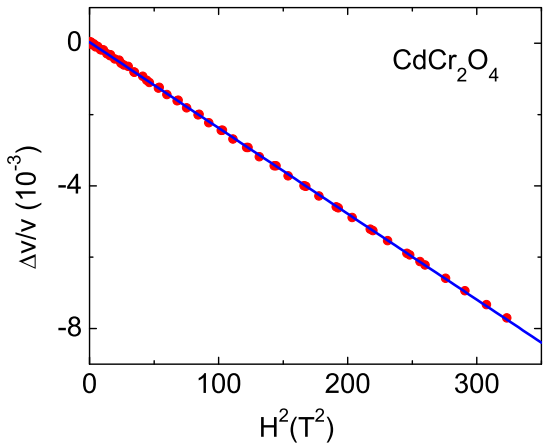}
\caption{(Color Online) A fit (solid line) of the velocity variation (symbols) with magnetic field from experiments at $T=8$ K.}
 \label{v-T-B-fit}
\end{figure}

We have already seen that the sound velocity depends on short-ranged spin correlations. The magnetic field affects the spin correlations, and, hence, the sound velocity is affected as well.  Ultrasound measurements as a function of magnetic field (at various temperatures) are presented in Fig. \ref{compare2}. Besides some low-field features (below $T_N$, which we will discuss later), the acoustic mode demonstrates a clear softening with increasing field. This general trend may be understood qualitatively as follows. In the presence of a magnetic field, the free energy of the system is given by
\begin{eqnarray}
\mathcal{F}=\mathcal{F}_0-\frac{1}{2}\chi(\epsilon)H^2,
\end{eqnarray}
where $\mathcal{F}_0$ is the free energy in the absence of the magnetic field and $\chi(\epsilon)$ is the magnetic susceptibility (which is a function of the strain field $\epsilon$ in the presence of magneto-elastic coupling). This gives an additional contribution to the elastic constants $c_{ij}=\frac{\partial^2\mathcal{F}}{\partial\epsilon_i\epsilon_j}$. For $\frac{\partial^2\chi}{\partial\epsilon_i\epsilon_j}\ne 0$,
\begin{eqnarray}
\Delta c_{ij}\propto H^2,
\end{eqnarray}
 and this is observed in the experiments (see Fig. \ref{v-T-B-fit}). A more microscopic consideration (for the paramagnetic phase) developed in Sec. \ref{sec_para_phase} suggests that the fractional change in the sound velocity in the presence of a magnetic field is given by
\begin{eqnarray}
\label{mag_velocity}
\left(\frac{\Delta v}{v}\right)_h=\frac{\alpha^2\delta^2J_0S^2}{18 M (v_{\bf \hat q}^0)^2}\left(\overline{\mathcal{E}}^{\mathrm{Tet}}_h-T_0\overline{C}_h^{\mathrm{Tet}}\right).
\end{eqnarray}
Here, $\overline{C}^{\mathrm{Tet}}_h$ and $\overline{\mathcal{E}}^{\mathrm{Tet}}_h$ are the {\em scaled} [see discussion following Eq. (\ref{final_exp}) and also Appendix \ref{appen2}] exchange parts (see below) of the specific heat and magnetic energy, respectively, {\em i.e.},
\begin{eqnarray}
\overline{C}^{\mathrm{Tet}}_h =2\overline{C}^{\mathrm{spin}}_h=\frac{2}{NK_BT_0^2}{\langle\langle (\sum_{\langle ij\rangle}{\bf S}_i\cdot{\bf S}_j)^2\rangle\rangle_h}
\end{eqnarray}
and
\begin{eqnarray}
\overline{\mathcal{E}}^{\mathrm{Tet}}_h = 2 \overline{\mathcal{E}}^{\mathrm{spin}}_h= \frac{2}{N} {\langle \sum_{\langle ij\rangle}{\bf S}_i\cdot{\bf S}_j\rangle_h}.
\end{eqnarray}
where, as before, the factor $2$ arises because each tetrahedron contains two spins. However, unlike the case of zero magnetic field, these quantities cannot be expressed (even within the single tetrahedron approximation) in closed analytic form. Hence, we resort to Monte Carlo simulations of a Heisenberg magnet on a single tetrahedron to calculate these expressions. We use a heat-bath Monte Carlo algorithm and find that we can fit the simulation data in the following phenomenological form:
\begin{eqnarray}
\left(\overline{\mathcal{E}}^{\mathrm{Tet}}_h-T_0\overline{C}_h^{\mathrm{Tet}}\right)=\phi(T_0)\left(\frac{\mu_BgH}{J_0S^2}\right)^2,
\label{monte_function}
\end{eqnarray}
where $\phi(T_0)$ is a function of only the scaled temperature introduced before [$T_0=T/(J_0S^2)$] and $H$ is the magnetic field measured in Tesla. Using this phenomenological form of the function [Eq. (\ref{monte_function})] in Eq. (\ref{mag_velocity}), we get
\begin{eqnarray}
\left(\frac{\Delta v}{v}\right)_h=4\frac{\alpha^2\delta^2J_0S^2}{18 M (v_{\hat{\bf q}}^0)^2}\phi(T_0)\left(\frac{\mu_BgH}{J_0S^2}\right)^2.
\label{field_fit_form}
\end{eqnarray}
This is the expression for the dependence of the sound velocity on the magnetic field in the cooperative paramagnetic regime. This expression is in accordance with our earlier expectations that $\Delta v/v\propto H^2$. While $\alpha$ determined from these data is broadly consistent with our earlier estimation, this agreement is no longer quantitative-- the data analyzed here are at temperatures too low for our perturbation theory. 
\subsection{Low-temperature ordered phase and role of the DM interaction}
\begin{figure}
\begin{center}
\vspace{-0.5cm}
\includegraphics[width=0.9\linewidth, keepaspectratio]{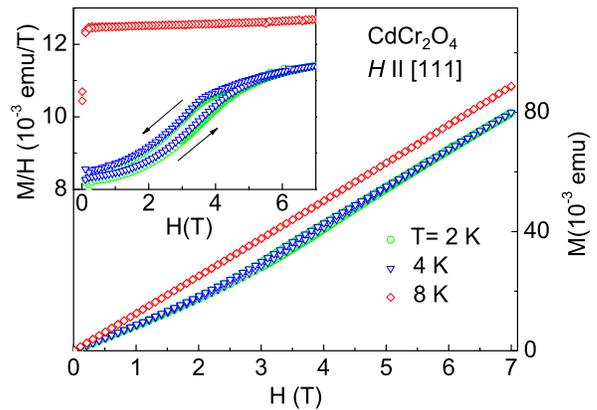}
\end{center}
\vspace{-0.5cm}
\caption{(Color online) Magnetization measured for magnetic fields applied along the [111] direction at temperatures of 2, 4, and 8 K. Zero-field cooled curves are shown. The inset shows the same data presented as $M/H$ vs $H$. The arrows indicate the field-sweep direction.}
\label{compare3}
\end{figure}

For temperatures below $T_{N}$, there is a kink like anomaly at about 4 T accompanied by hysteresis, see Fig. \ref{compare2}. Magnetization measurements in this temperature and magnetic-field region (Fig. \ref{compare3}) also exhibit a similar hysteretic behavior in the temperature range below $T_{N}$ with an anomaly at a magnetic field of about 4 T. This is clearly visible for the $M/H$ vs $H$ data shown in the inset of Fig. \ref{compare3}. Above $T_{N}$, the magnetization exhibits a linear magnetic-field dependence.

Chern {\em et al.} {\cite{2006_chern}} showed that the long-ranged spiral results from weak DM interactions. Their presence ($H_{DM}=\sum_{\langle ij\rangle}{\bf D_{ij}\cdot S_i\times S_j}$, with non-collinear ${\bf D_{ij}}$) breaks the global spin-rotation symmetry to discrete lattice symmetries and results in an energy gap (per tetrahedron) of the order of $\Delta\approx 0.7\ K$ (obtained from ab initio calculations{\cite{2006_chern}}). The Zeeman term, on the other hand, favours a four-sublattice canted spin structure. A first-order transition between the spiral and the canted states \cite{2006_kimura,2007_matsuda2} is responsible for the observed low-temperature anomalies. An estimate of the magnitude of the magnetic field at which this transition occurs may be obtained by comparing the energies of the spiral and the canted states. The Zeeman-energy gain (per tetrahedron) in the canted state is given by $E_T=a H^2$ with $a=-0.0431\ (K/T^2)$ and $H$ in Tesla. {\cite{2010_bhattacharjee}} Thus, a rough estimate of the magnitude of the magnetic field at which the transition occurs is $H_c\approx 4.3\ T$. (We have assumed the case of a strongly first-order transition where the undistorted spiral changes suddenly to the canted state at $H=H_c$. The hysteresis in the sound-velocity and magnetization data justifies our assumption.)

Below $H_c$, in the spiral state, the spin correlations do not change appreciably with the magnetic field leading to an almost constant sound velocity, as seen in experiment(Fig. \ref{compare2}). Above ($H_c$), in the canted state, the spin correlations, however, are sensitive to the magnetic field, and the sound velocity changes with magnetic field $(\Delta v/v\propto H^2)$. Our estimate of $H_c$ matches fairly well with the value of the field at which anomalies are seen in the experiments (Figs. \ref{compare2} and \ref{compare3}). But it is not clear why the characteristic field of 4 T is somewhat lower than 5.7 T, where similar anomalies have been reported in ESR measurements. \cite{2006_kimura}
\begin{figure}
\begin{center}
\vspace{-0.5cm}
\includegraphics[width=0.9\linewidth, keepaspectratio]{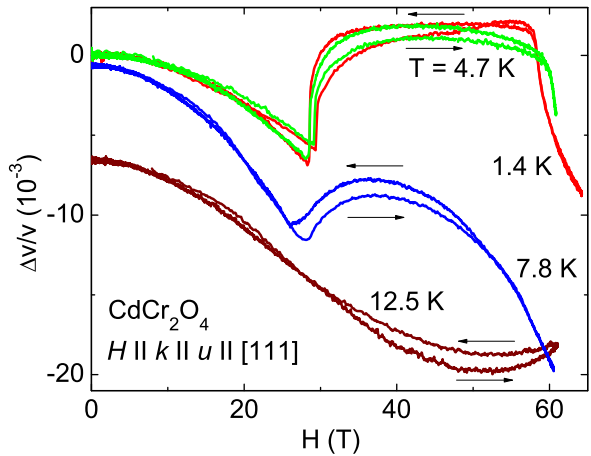}
\end{center}
\vspace{-0.5cm}
\caption{(Color online) Change of the sound velocity for the $c_{L}$ mode in CdCr$_2$O$_4$ measured in pulsed magnetic fields at different temperatures and an ultrasound frequency of 81 MHz. The experimental geometry is $H \parallel k \parallel u \parallel [111]$.}
\label{compare4}
\end{figure}
\subsection{High magnetic field regime}

We have also performed pulsed magnetic-field experiments in CdCr$_2$O$_4$ in order to study the acoustic behavior in the region of the magnetization plateau. Results for the sound velocity in magnetic fields up to 63 T at temperatures of 1.4, 4.7, 7.8, and 12.5 K are shown in Fig. \ref{compare4}. At 1.4 and 4.7 K, the sound velocity decreases first with the characteristic $H^2$ behavior, then demonstrates a jump at the magnetic field where the magnetization plateau appears. This anomaly corresponds to a first-order phase transition from the four-sublattice canted spin structure to a collinear spin configuration with three spins up and one spin down at each Cr$^{3+}$ tetrahedra. Previously, a large magnetostriction has been reported at this phase transition. \cite{2005_ueda} A cubic crystallographic structure has been suggested from high-field x-ray experiments at the plateau state. \cite{2006_inami} Recent elastic-neutron-scattering experiments in pulsed magnetic fields \cite{2010_matsuda} showed that the magnetic structure at the half-magnetization plateau phase has a cubic $P4_332$ symmetry. It has been proposed \cite{2004_penc,2007_matsuda2} that the lattice distortion stabilizes the three-up one-down collinear spin configuration. Once again, the lattice distortion complicates the theoretical analysis of this regime. Indeed, there is only a slight change of the sound velocity within the magnetization-plateau range. This is because the spin correlations are locked at fixed values within this plateau. The plateau terminates at approximately 58 T, confirmed by a sharp anomaly, i.e., an abrupt decrease in the sound velocity. This anomaly corresponds to a phase transition to a non-collinear canted spin configuration. \cite{2008_kojima, 2004_penc} It is interesting to note that the hysteresis in the sound velocity not only takes place at the first-order phase transition around 28 T, but also spreads along the whole plateau range up to 58 T, showing a complicated interplay between the spin and the lattice degrees of freedom within the magnetization plateau. The sound-velocity change, which takes place between 58 and 63 T, is even larger than the anomaly at 28 T. The highest applied magnetic field of 63 T is not sufficient to detect the complete sound-velocity change. No hysteresis has been detected at 58 T, pointing to a second-order type of this phase transition. It is worth noting that the magnetization only exhibits a smooth kink like anomaly at this phase transition. \cite{2008_kojima} A transverse spin order, which is equivalent to a Bose-Einstein condensation of magnons, is predicted from quantum-fluctuation theory at magnetic fields just above the plateau. \cite{2008_bergman}

The first-order phase transition at about 28 T could be resolved by magnetization measurements up to temperatures slightly above $T_{N}$. \cite{2005_ueda} Our pulsed-field ultrasound measurements performed at 7.8 K (Fig. \ref{compare4}) also clearly reveal an anomaly in the sound velocity at this field. The anomaly at 28 T is somewhat smoother than at lower temperatures but still clearly evident. The hysteresis survives, and the total change in the sound velocity is even larger at higher temperatures approaching 2 \% between 0 and 60 T. Further temperature increase suppresses the anomaly at 28 T leading to a broad minimum in the sound velocity at about 50 T (see the lower curve in Fig. \ref{compare4}). Note that the hysteresis is still observable at 12.5 K.
\section{Conclusion}

\label{sec_conclusion}

In this paper, we have successfully characterized the strength of the magneto-elastic coupling in CdCr$_2$O$_4$ and have been able to explain the general features except at highest fields. We would like to mention that, in the current paper, we deal with the acoustic $c_L$ mode where various deformations are involved (see the definition of $c_L$ mode above). This fact complicates the symmetry analysis of the obtained data. In addition, below $T_{N}$, in the tetragonal phase, there are three types of domains corresponding to an elongation of the $c$ axis, and all of them contribute to the acoustic $c_L$ mode. In this regard analysis of the sound mode along one of the axes of the cubic/tetragonal crystal may yield more theoretically tractable results. A detailed analysis of these other modes and their relation to the different elastic constants are given in Ref. \onlinecite{2010_bhattacharjee}.

We also note that hydrodynamic calculations suggest that, in the regime $\vert\Theta_{CW}\vert\ll T\ll T_M$ (where $T_M$ is the melting temperature of the crystal), $\Delta v/v\rightarrow K$, where $K\neq 0$ is a constant. Thus, in this limit, there is a constant shift in the sound velocity compared to an iso-structural compound without spin-phonon coupling. This result is rather interesting and requires further investigation.

In general, the sound waves, being hydrodynamic modes, are very robust, and they couple to a large number of low-lying excitations. It is relatively straightforward to measure the sound velocity with high enough accuracy, and this is especially true for the present case of magnetoelastic transitions. However, since they themselves do not go critical and also because they generally interact with most low-energy modes (mentioned above) disentangling microscopic information from them is less straightforward. 

To summarize, we have presented a magneto-acoustic study of the frustrated spin system CdCr$_2$O$_4$. Strong sound-velocity anomalies have been observed at the magnetic phase transitions in CdCr$_2$O$_4$. We have been able to characterize the spin-strain coupling, which is crucial and determines the underlying physics of this compound, reducing the geometric frustration effect and lifting the degeneracy in the system. The detailed comparison of experiment and theory in different regimes suggests that the dominant part of the variation of the sound velocity at low temperatures is due to the spin-lattice interaction, which in turn, may be modeled as an exchange-striction phenomenon.
\vspace{-0.5cm}
\begin{acknowledgements}
This research was partly supported by EuroMagNET II under EU Contract No. 228043, DFG Grant No. LE 967/6-1, ESF-HFM and the DFG via Transregional Collaborative Research Center TRR 80 (Augsburg  Munich). S.B. and M.E.Z. acknowledge the Visitors Program of the Max Planck Institute for Complex Systems (MPI-PKS) for hospitality.

\end{acknowledgements}
\appendix
\section{}
\label{appen}
In this appendix, we give an outline of the perturbative calculation for both the high-temperature paramagnetic phase as well as the low-temperature ordered phase. 
\vspace{-0.5cm}
\subsection{High-temperature paramagnetic phase}
\label{appen1}
Here, we derive Eq. (\ref{semifinal_exp}) from Eq. (\ref{delta_v_expression}) for the sound velocity in the $[111]$ direction. From Eqs. (\ref{eff_potential}) and (\ref{self_energy}), we find that the phonon self-energy is given by
\begin{eqnarray}
\Sigma(\imath\Omega_n,{\bf q})=V_{\bf q q}=\langle U^{(2)}_{\bf q q}\rangle-\beta\langle\langle U^{(1)}_{\bf q}U^{(1)}_{-\bf q}\rangle\rangle,
\end{eqnarray}
where, from Eqs. (\ref{potential1}) and (\ref{potential2}), we have
\begin{widetext}
\begin{eqnarray}
\nonumber
\langle\langle U^{(1)}_{\bf q}U^{(1)}_{-\bf q}\rangle\rangle&=&\frac{-\beta}{2MN\omega_{\bf q}^0}\sum_{\langle ij\rangle}\sum_{\langle lm\rangle}\langle\langle\left({\bf S}_i\cdot{\bf S}_j\right)\left({\bf S}_l\cdot{\bf S}_m\right)\rangle\rangle (e^{\imath {\bf q}\cdot{\bf r}_i}-e^{\imath {\bf q}\cdot{\bf r}_j}) \left[\frac{\partial J_{ij}}{\partial{\bf R}_{ij}}\cdot {\bf e}_{\bf q}\right]\left[\frac{\partial J_{lm}}{\partial{\bf R}_{lm}}\cdot {\bf e}_{-{\bf q}}\right] (e^{-\imath {\bf q}\cdot{\bf r}_l}-e^{-\imath {\bf q}\cdot{\bf r}_m}),\\
\\
\langle U^{(2)}_{\bf q q}\rangle&=&\frac{1}{2\ M N}\frac{1}{\omega_q^0}\sum_{\langle ij\rangle}\langle{\bf S}_i\cdot{\bf S}_j\rangle\left( e^{-\imath{\bf q}\cdot{\bf r}_j}-e^{-\imath{\bf q}\cdot{\bf r}_i}\right)\left({\bf e}_{-\bf q}\cdot\frac{\partial^2 J_{ij}}{\partial{\bf R}_i\partial{\bf R}_j}\cdot{\bf e}_{\bf q}\right)\left( e^{\imath{\bf q}\cdot{\bf r}_i}-e^{\imath{\bf q}\cdot{\bf r}_j}\right).
\end{eqnarray}
Taking the limit given by Eq. (\ref{delta_v_expression}) we have (for longitudinally polarized phonons)
\begin{eqnarray}
\frac{\Delta v}{v}=\frac{1}{2MN(v^0_{\bf \hat q})^2}\left[\left(\frac{\delta^2}{J_0}\frac{\partial^2 J}{\partial \delta^2}\right)\sum_{\langle ij\rangle} ({\hat{\bf\delta}_{ij}\cdot\hat{\bf q}})^4\langle J_0{\bf S}_i\cdot{\bf S}_j\rangle
-\beta \left(\frac{\delta}{J_0}\frac{\partial J}{\partial\delta}\right)^2\left<\left<\left(\sum_{\langle ij\rangle}\left({\hat{\bf\delta}_{ij}\cdot\hat{\bf q}}\right)^2 J_0{\bf S}_i\cdot{\bf S}_j\right)^2\right>\right>\right],
\end{eqnarray}
where, $\delta$ is the distance between two neighbouring Cr spins, which is same for all directions (${\bf\hat\delta}_{ij}$). In a paramagnetic state and a general direction of the sound velocity $(\hat q)$, we have 
\begin{eqnarray}
\frac{\Delta v}{v}=\frac{1}{2M(v^0_{\bf \hat q})^2}\left[\left(\frac{\delta^2}{J_0}\frac{\partial^2 J}{\partial \delta^2}\right)\mathcal{A}(\hat{\bf q})\mathcal{E}^{\mathrm{spin}}-\left(\frac{\delta}{J_0}\frac{\partial J}{\partial\delta}\right)^2\mathcal{B}(\hat{\bf q})C^{\mathrm{spin}}T\right].
\end{eqnarray}
\end{widetext}
In the last expression, we have used
\begin{eqnarray}
\mathcal{E}^{\mathrm{spin}}&=&\frac{1}{N}\langle J_0\sum_{\langle ij\rangle}{\bf S}_i\cdot{\bf S}_j\rangle,\\
K_BT^2C^{\mathrm{spin}}&=&\frac{1}{N}\left<\left<\left(J_0\sum_{\langle ij\rangle}{\bf S}_i\cdot{\bf S}_j\rangle\right)^2\right>\right>.
\end{eqnarray}
For ${\bf{\hat q}}=[111]$, we have (from the fact that ${\bf {\hat\delta}}\cdot{\bf {\hat q}}=\sqrt{\frac{2}{3}}$ for exactly half of the bonds),
\begin{eqnarray}
\mathcal{A}({\bf{\hat q}})=\mathcal{B}({\bf{\hat q}})=\frac{2}{9}.
\end{eqnarray}
Hence, we get Eq. (\ref{semifinal_exp}).
\subsection{Low-temperature magnetically ordered phase}
\label{appen1b}
Here, we outline the calculation for the fractional change of sound velocity in the low-temperature ordered state due to magnon-phonon scattering. We assume a collinear two sub-lattice Neel order. While the differences in the lattice structure may lead to variations in different non-universal pre factors, the temperature dependence does not change as long as the magnon dispersion remains linear. Similar calculations (not shown) can be performed for the ferromagnetic case as well. We start by defining two species of HP boson operators,
\begin{eqnarray}
\nonumber
Sublattice\ 1&:& S^\dagger=\sqrt{2S}a,S^-=\sqrt{2S}a^\dagger,S^z=S-a^\dagger a,\\
\nonumber
Sublattice\ 2&:& S^\dagger=\sqrt{2S}c,S^-=\sqrt{2S}c^\dagger,S^z=S-c^\dagger c.\\
\end{eqnarray}
The Hamiltonian is given by
\begin{eqnarray}
H=H_P+H_{sp}+H_1+H_2,
\end{eqnarray}
where
\begin{eqnarray}
H_P&=&\hbar\sum_{\bf k}\omega^0_{\bf k}b^\dagger_{\bf k}b_{\bf k},\\
H_{sp}&=&-\frac{J_0S^2zN}{2}-\sum_{\bf k}\omega_{\bf k}^s\left(\alpha_{\bf k}^\dagger \alpha_{\bf k}+\beta_{\bf k}^\dagger\beta_{\bf k}\right).
\end{eqnarray}
($z$ is the coordination number; $\omega^s_{\bf k}=c_sk$ is the AF magnon frequency.) $\alpha$ and $\beta$ are the Bogoliubov rotated bosonic operators given by
\begin{eqnarray}
\left[\begin{array}{c}
a_{\bf k}\\
b^\dagger_{-\bf k}\\
\end{array}\right]=
\left[\begin{array}{cc}
\cosh{\theta_{\bf k}} & \sinh{\theta_{\bf k}}\\
\sinh{\theta_{\bf k}} & \cosh{\theta_{\bf k}}\\
\end{array}\right]
\left[\begin{array}{c}
\alpha_{\bf k}\\
\beta^\dagger_{-\bf k}\\
\end{array}\right], 
\end{eqnarray}
and 
\begin{eqnarray}
\tanh{2\theta_{\bf k}}=-\frac{\sum_{\bf \delta}e^{\imath {\bf k\cdot\delta}}}{z}.
\end{eqnarray}
Also 
\begin{eqnarray}
H_1&=&\sum_{\bf k,q}A_{\bf k}\Psi^\dagger_{\bf k+q}\cdot\mathcal{M}_{\bf k,q}\cdot \Psi_{\bf q},\\
\nonumber
H_2&=&\frac{1}{2}\sum_{\bf k}\Gamma^{(0)}_{\bf k}A_{\bf k}A_{\bf -k}\\
&+&\frac{1}{2}\sum_{\bf k,k',q}A_{\bf k}A_{\bf -k'}\Psi^\dagger_{\bf k-k'+q}\cdot\mathcal{N}_{\bf k,k',q}\cdot\Psi_{\bf q},
\end{eqnarray}
where $\Psi_{\bf k}^\dagger=[\alpha^\dagger_{\bf k},\beta_{-\bf k}]$, $A_{\bf k}=b_{\bf k}+b^\dagger_{-\bf k}$ is the phonon displacement operator and 
\begin{eqnarray}
\nonumber
\Gamma^{(0)}_{\bf k}=-\frac{\alpha^2J_0S^2\hbar}{4M\omega^0_{\bf k}}\sum_{\delta}\left(1-e^{\imath{\bf k\cdot\delta}}\right)\left(1-e^{-\imath{\bf k\cdot\delta}}\right)({\bf \hat\delta\cdot\hat e_{k}})({\bf \hat\delta\cdot\hat e_{-k}})\\
\end{eqnarray}
and $\mathcal{M}_{\bf k,q}$ and $\mathcal{N}_{\bf k,k',q}$ are $2\times 2$ matrices
\begin{eqnarray}
\mathcal{M}_{\bf k,q}&=&\mathcal{C}_{\bf k+q}\cdot\Delta_{\bf {k,q}}\cdot\mathcal{C}_{\bf q},\\
\mathcal{N}_{\bf k,k',q}&=&\mathcal{C}_{\bf k-k'+q}\cdot\mathcal{O}_{\bf k,k',q}\cdot\mathcal{C}_{\bf q},
\end{eqnarray}
where,
\begin{eqnarray}
\mathcal{C}_{\bf q}=\left(\begin{array}{cc}
\cosh{\theta_{\bf k}} & \sinh{\theta_{\bf k}}\\
\sinh{\theta_{\bf k}} & \cosh{\theta_{\bf k}}\\
\end{array}\right)
\end{eqnarray}
and
\begin{widetext}
\begin{eqnarray} 
\Delta_{\bf k,q}&=&-\frac{J_0S\alpha}{2}\sqrt{\frac{\hbar}{2MN\omega^0_{\bf k}}}\sum_{\bf \delta}\left(\left(1-e^{\imath {\bf k\cdot\delta}}\right)({\bf \hat\delta\cdot\hat e_{k}})\left[\begin{array}{cc}
1 & e^{\imath{\bf q\cdot\delta}}\\
e^{\imath({\bf k+q}){\cdot{\bf \delta}}} & 1\\
\end{array}\right]\right),\\
\mathcal{O}_{\bf k,k',q}&=&\frac{\alpha^2J_0S\hbar}{4MN\sqrt{\omega^0_{\bf k}\omega^0_{\bf k'}}}\sum_{\bf \delta}\left(\left(1-e^{\imath{\bf k\cdot\delta}}\right)\left(1-e^{-\imath{\bf k'\cdot\delta}}\right)({\bf \hat\delta\cdot\hat e_{k}})({\bf \hat\delta\cdot\hat e_{-k'}})\left[\begin{array}{cc}
1 & e^{\imath{\bf q\cdot\delta}}\\
e^{\imath({\bf k-k'+q})\cdot{\bf\delta}}\\
\end{array}\right]
\right).
\end{eqnarray}
\end{widetext}
The phonon Green's function is given by Eq. (\ref{matsubara_green}) and the self-energy consists of two parts 
\begin{eqnarray}
\Sigma({\bf q},\imath\Omega_n)=\Sigma_1({\bf q},\imath\Omega_n)+\Sigma_2({\bf q},\imath\Omega_n).
\end{eqnarray}
Defining the bare matrix Green's function for the magnons as $G_m^{ij}({\bf k},\tau)=-\langle\mathcal{T}\{\Psi^i_{\bf k}(\tau){\psi^j_{\bf k}}^\dagger(0)\}\rangle$ or its Fourier transform,
\begin{eqnarray}
G_m({\bf k},\imath\Omega_n)=\left(\begin{array}{cc}
\frac{1}{\imath\Omega_n-\omega_{\bf k}^s} & 0\\
0 & \frac{-1}{\imath\Omega_n+\omega_{\bf k}^s}\\
\end{array}\right),
\end{eqnarray}
we find that the contributions to the self-energy from $H_1$ and $H_2$ are given by
\begin{widetext}
\begin{eqnarray}
\label{exp11}
\Sigma_1(\imath\Omega_n,{\bf q})&=&-\frac{1}{\beta}\sum_{{\bf q'},\eta_n}\mathcal{M}_{\bf -q,q+q'}^{ij}\mathcal{M}_{\bf q,q'}^{ji}G^{ii}_m({\bf q'},\imath\eta_n)G^{jj}_m({\bf q+q'},\imath\Omega_n+\imath\eta_n),\\
\label{exp12}
\Sigma_2(\imath\Omega_n,{\bf q})&=&\Gamma^{(0)}_{\bf q}+\frac{1}{2}\sum_{\bf q'}\left(\mathcal{N}^{ii}_{\bf q,q,q'}+\mathcal{N}^{ii}_{\bf -q,-q,q'}\right)G_m^{ii}(\imath0^-,{\bf q'}).
\end{eqnarray}
\end{widetext}
Now, the frequency and the fractional change of sound velocity are given by Eqs. (\ref{real_self_energy}) and (\ref{delta_v_expression}). In general, the integrals in Eqs. (\ref{exp11}) and (\ref{exp12}) are hard to calculate for a general direction of $\hat{\bf q}$. While this can be done numerically, here, we only concentrate on their dependence on temperature, which gives the result given by Eq. (\ref{low_t}).
\section{}
\label{appen2}
Here we give the explicit expressions for $\overline{\mathcal{E}}^{\mathrm{Tet}}$ and $\overline{C}^{\mathrm{Tet}}$, which are the energy and specific heat of nearest-neighbour classical Heisenberg AF (with spins of unit magnitude) on the corners of a tetrahedron. The temperature is scaled by $1/(J_0S^2)$, {\em i.e.}, $T_0=T/J_0S^2$, to compare with experiments.
\begin{widetext}
$\overline{\mathcal{E}}^{\mathrm{Tet}}$ and $\overline{C}^{\mathrm{Tet}}$ in Eq. (\ref{scaled_final_exp}) can be obtained from Ref. \onlinecite{1999_moessner} and are given by
\begin{eqnarray}
\overline{\mathcal{E}}^{\mathrm{Tet}}&=&-2+\frac{3}{2}T_0-\frac{e^{2/T_0}T_0^3}{\overline{\mathcal{Z}}^{\mathrm{Tet}}\sqrt{32\pi}}\left(3-4 e^{-2/T_0}+e^{-8/T_0}\right),\\
\overline{C}^{\mathrm{Tet}}&=& \frac{3}{2}+\frac{T_0e^{2/T_0}}{\overline{\mathcal{Z}}^{\mathrm{Tet}}\sqrt{32\pi}}\left[\left(\overline{\mathcal{E}}^{\mathrm{Tet}}-T_0\right)\left(3-4e^{-2/T_0}+e^{-8/T_0}\right)+6\left(1-e^{-8/T_0}\right)\right],
\end{eqnarray}
where $\overline{\mathcal{Z}}^{\mathrm{Tet}}$ is the {\em scaled} partition function of the tetrahedron, which is given by \cite{1999_moessner}
\begin{eqnarray}
\overline{\mathcal{Z}}^{\mathrm{Tet}}=T_0^{3/2}e^{2/T_0}\left[2 \mathrm{Erf}\left({\sqrt{2/T_0}}\right)-\mathrm{Erf}\left(\sqrt{8/T_0}\right)-\sqrt{\frac{T_0}{8\pi}}\left(1-e^{-2/T_0}\right)^2\left(3-2 e^{-2/T_0}+e^{-4/T_0}\right)\right].
\end{eqnarray}
Here, $\mathrm{Erf}(x)$ is the error function defined as $Erf(x)=\sqrt{4/\pi}\int_0^xe^{-t^2}dt$.
\end{widetext}

\end{document}